# Volatile Sample Return in the Solar System

Stefanie N. Milam[1], NASA GSFC, 301.614.6902, stefanie.n.milam@nasa.gov

Jason P. Dworkin[1], Jamie E. Elsila[1], Daniel P. Glavin[1], Perry A. Gerakines[1], Julie L. Mitchell[2], Keiko Nakamura-Messenger[2], Marc Neveu[1,4], Larry Nittler[3], James Parker[1], Elisa Quintana[1], Scott A. Sandford[5], Joshua E. Schlieder[1], Rhonda Stroud[6], Melissa G. Trainer[1], Meenakshi Wadhwa[7], Andrew J. Westphal[8], Michael Zolensky[2], Dennis Bodewits[9], Simon Clemett[2]

[1]NASA Goddard Space Flight Center
[2]NASA Johnson Space Center
[3]Carnegie Institute of Washington
[4]University of Maryland, College Park
[5]NASA Ames Research Center
[6]Naval Research Laboratory
[7]Arizona State University
[8]UC Berkeley Space Sciences Laboratory
[9]Auburn University

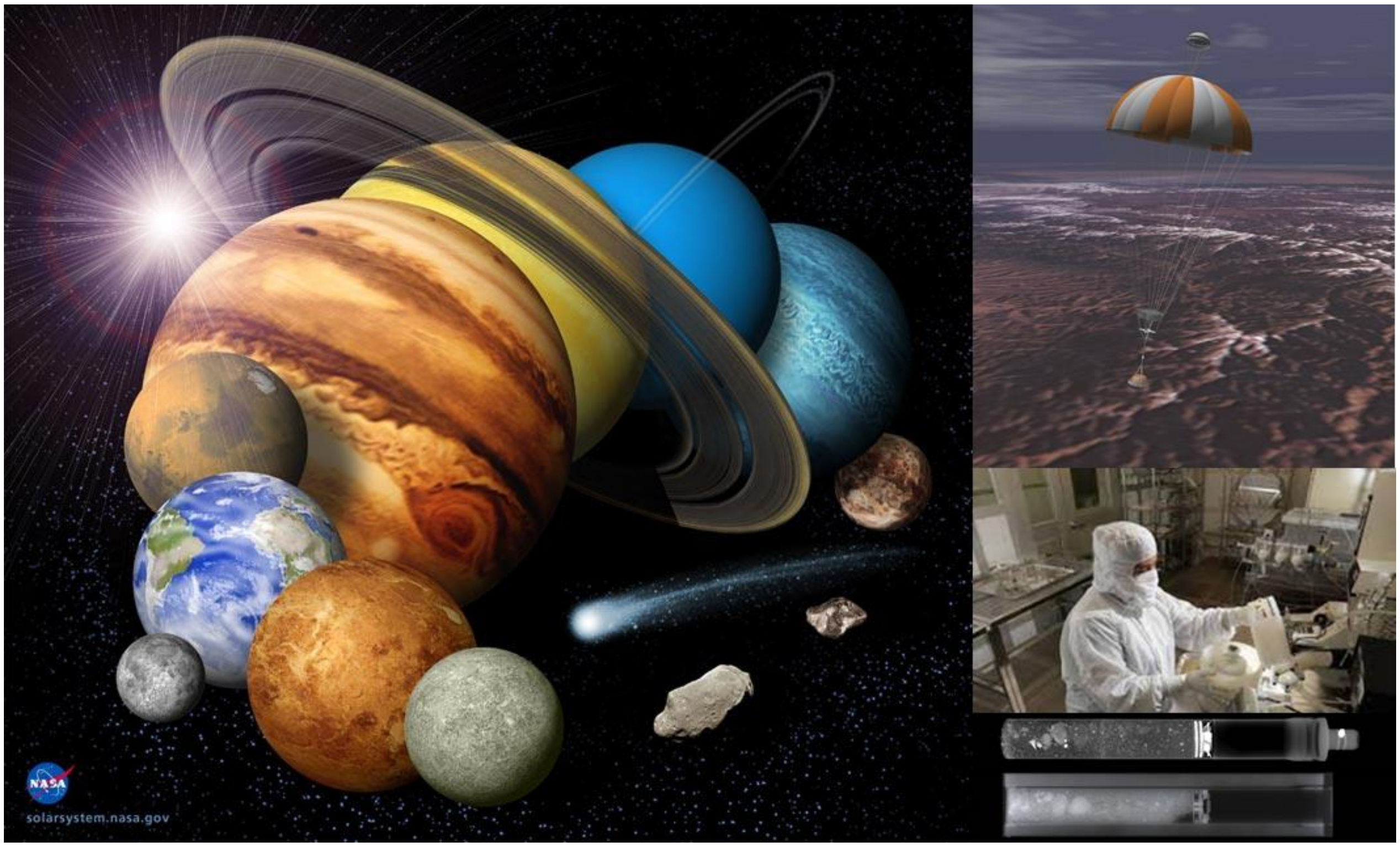

Contributions/Endorsements: D. Bockelee-Morvan (Paris Observatory), B. Cohen (NASA/GSFC), T. Barrett (Open University)

**Abstract:** We advocate for the realization of volatile sample return from various destinations including: small bodies, the Moon, Mars, ocean worlds/satellites, and plumes. As part of recent mission studies (e.g., Comet Astrobiology Exploration SAmple Return (CAESAR) and Mars Sample Return), new concepts, technologies, and protocols have been considered for specific environments and cost. Here we provide a plan for volatile sample collection and identify the associated challenges with the environment, transit/storage, Earth re-entry, and curation. Laboratory and theoretical simulations are proposed to verify sample integrity during each mission phase. Sample collection mechanisms are evaluated for a given environment with consideration for alteration. Transport and curation are essential for sample return to maximize the science investment and ensure pristine samples for analysis upon return and after years of preservation. All aspects of a volatile sample return mission are driven by the science motivation: isotope fractionation, noble gases, organics and prebiotic species; plus planetary protection considerations for collection and for the sample.

- The science value of sample return missions has been clearly demonstrated by previous sample return programs and missions.
- Sample return of volatile material is key to understanding (exo)planet formation, evolution, and habitability.
- Returning planetary volatiles poses unique and potentially severe technical challenges. These include preventing changes to samples between (and including) collection and analyses, and meeting planetary protection requirements.

## 1. The scientific value of volatile sample return:

Volatiles –defined here as elements or compounds that are gases or liquids at standard state– play critical roles in wide swaths of planetary science. The distributions of volatiles across the Solar System are important for informing planet formation and migration theory, understanding planetary evolution, and interpreting observations of exoplanets. The presence of water and organic molecules is critical for the emergence of life and the habitability of planetary environments. Ices in comets, asteroids, moons, and outer Solar-System bodies inform the Solar System's starting (interstellar) materials, protoplanetary disk processes, geological and geochemical processes, interactions with the space environment, and the types of volatiles delivered to the terrestrial planets early in Solar System history. Atmospheric compositions inform planetary origins and geologic evolution. Although some information on planetary volatiles can be obtained from a highly sensitive and precise analysis of meteorites; alteration, exchange, and lack of context mean we are forced to rely on less sensitive measurements through telescopic and *in situ* spacecraft exploration (Fig 1).

Sensitivity, context, and cleanliness are enabled by sample return missions. Returning samples allows for the use of state-of-the-art analyses, providing for the ultimate current precision, sensitivity, resolution, and reliability, thus avoiding limitations associated with cost, power, mass, and reliability that would have affected or precluded making similar measurements *in situ*. As a result, the mission's payload effectively includes all the world's current and future analytical instrumentation. Since returned samples come from known bodies, they can be placed into geologic context and provide crucial complementary information with other studies of the parent object. Once returned to Earth, they are maintained in controlled curatorial environments that allow the samples to be used for current and future studies. Analyses of the returned samples can be iterative and fully adaptive — results are not limited by the selection of the initial spacecraft's design or by ideas current at the time of the flight of the spacecraft.

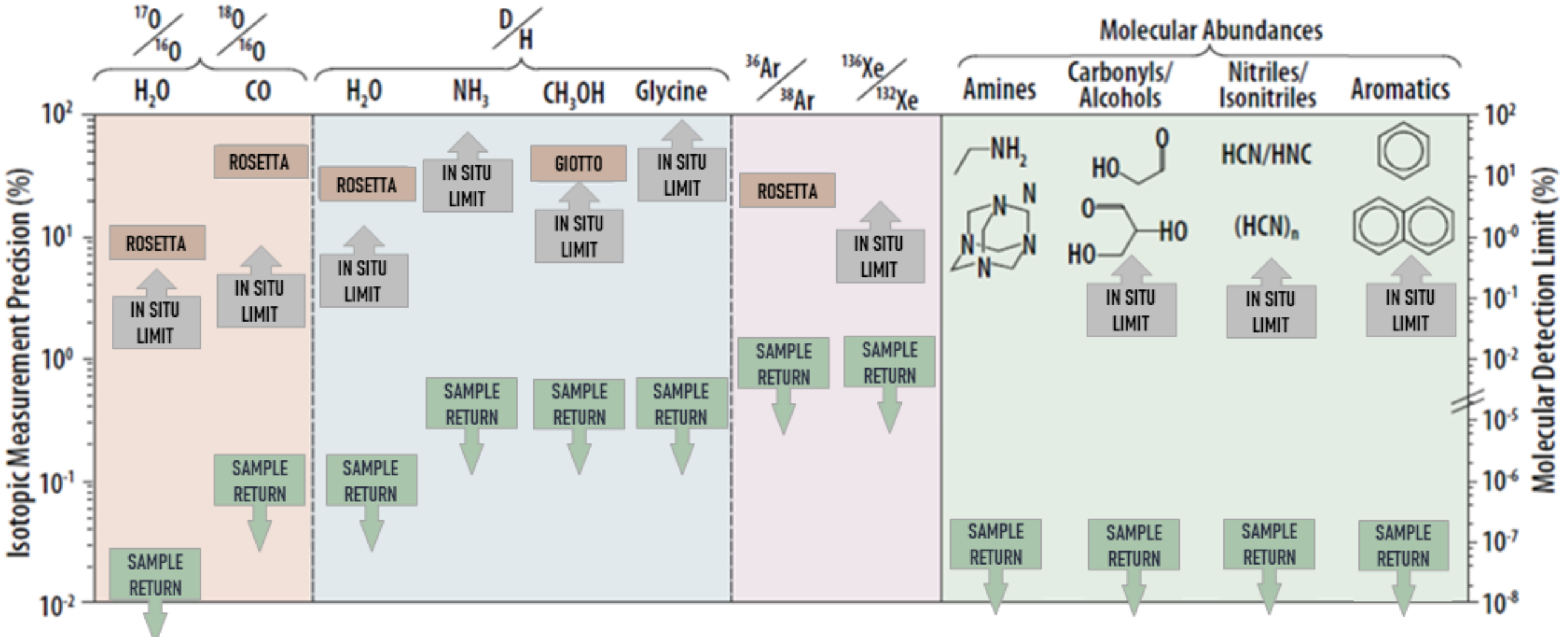

*Fig. 1. Volatile measurement capabilities for returned samples are superior to those for* in situ *spacecraft. Here examples for a comet is shown Left: isotopic measurement precision for selected volatiles. Right: molecular detections for selected classes of organics. Data from [1].*

The value of sample return missions has been clearly demonstrated by previous sample return programs and missions, including *Apollo, Luna, Stardust, Hayabusa,* and *Genesis,* all of which profoundly increased our understanding of key scientific questions in planetary science. However, except for *Genesis*, which returned atoms of the solar wind, all of these returned rocky material containing at best a tiny fraction of volatile compounds. Returning volatiles from planetary bodies poses unique and potentially severe technical challenges depending on the temperature and pressure requirements of the analytes (Fig 2).

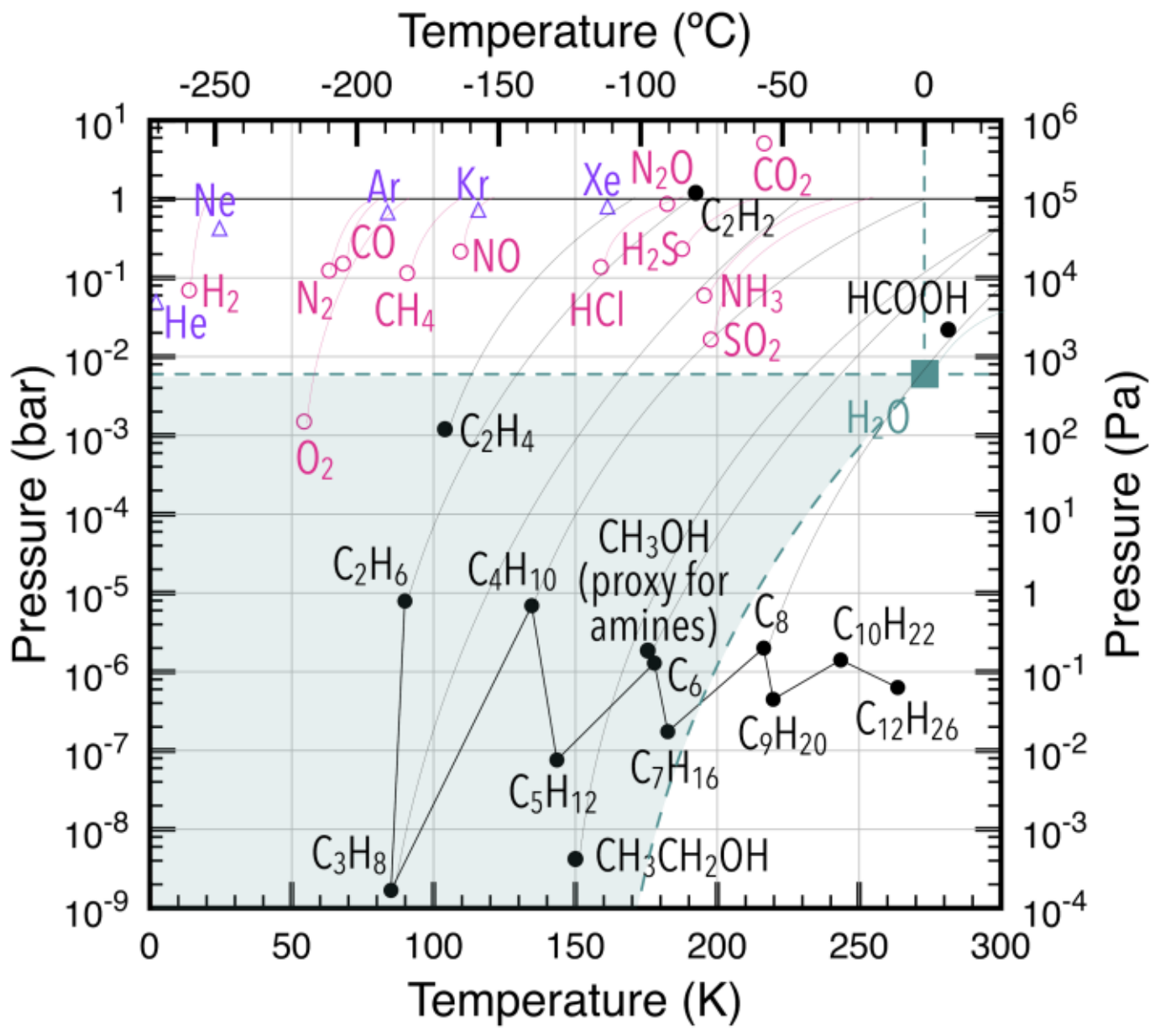

*Fig. 2. The maximum temperature of a sample for return is related to the triple point (water, noble gases, simple organics, and other small molecules) [2]. Melting temperatures are close to triple point temperatures, but can be lower in mixtures. Triple points are joined to the 1 bar horizontal line by vapor pressure curves. Methanol is a good proxy for amines and alkanes a good proxy for organics >C10. No aqueous reactions occur in the shaded zone.*

## 2. Rationales and Challenges for Sample Return Destinations:

***Icy Surfaces.*** Although refractory material has been returned from asteroids, sampling volatile material may only be possible at worlds large and cold enough to have retained their surface volatiles. The permanently-shadowed craters of the Moon may contain a record of solar system volatile reservoirs but they are extremely challenging environments in which to operate. *Artemis* may have the capacity to return volatiles from the south pole of the **Moon** but the capacity for volatile preservation and curation must be incorporated into the architecture. Lunar sample return specifications are being defined elsewhere [3]. **Ceres** is uniquely similar to both carbonaceous asteroids and ocean worlds. Ceres is blanketed by C- and N-bearing material altered by liquid water [4] with regions that contain exposed ice [5] or abundant aliphatic organic compounds [6]. Ceres has hosted globally an abiotic organic chemistry akin to that of carbonaceous meteorites and maybe an even more advanced chemical evolution [7]. Only sample return from its near-surface would enable measurements of the distribution of volatiles and organics and their isotopic and enantiomeric compositions to elucidate their synthesis pathways (Fig. 1; see also Ceres PMCS report). The science is just as intriguing for sample return from a **comet**. We have a glimpse of the rich chemistry, dynamics, and the natal composition of the Solar System from the refractories returned by *Stardust*. Volatile sample return, however, is needed for detailed studies of the organic composition within the icy body to understand the volatile isotopic composition preserved in the nucleus and the molecular complexity present prior to sublimation or processing. Sampling requires increased operational complexity and may require a landing as opposed to a touch-and-go to avoid thermally altering and contaminating the sample with volatiles during collection (see white paper by Nakamura-Messenger et al.).

***Plumes*** sample fresh interior volatiles from **Io**, **Enceladus**, **Triton**, and probably **Europa**. Using a 1 $m^2$ collector would allow retrieval of 40 mg at Io [8], 0.5 – 6 mg at Enceladus [9], and up to 1 g at Triton [10], which are comparable to Stardust's and Genesis' ~10 $mg/m^2$. Collection challenges include intermittency of eruptions (Io, Europa, and Triton) as well as Jupiter's radiation and a long return time (e.g., >13 years at Enceladus [11,12]).

***Dense Atmospheres.*** Venus and Titan have atmospheres denser than Earth and are both high-priority destinations for surface and atmospheric investigations to untangle the history of Solar System evolution and habitability. Atmospheric sample return even from a high atmosphere fly-through of Titan or Venus would provide valuable samples. The abundances and isotopic compositions of noble gases are needed to elucidate not only the history of how **Venus** evolved to its current state, but also to give the evolutionary context of Mars and Earth [13]. High-precision isotopic studies of the condensed-phase H, S, and O species would provide insight into the history of volcanism and atmospheric oxidation state to constrain the mixing between the lower and upper atmosphere of Venus and the S cycle. While much can be accomplished *in situ* [14], an atmospheric sample returned from below the homopause would permit more detailed analyses of low-abundance isotopes such as from Xe and O, provided sample integrity from collection is understood [15]. If a surface sample is collected [17] an atmospheric sample should also be returned. **Titan** atmospheric sample return would also give needed noble gas isotopic information, as predicted Xe abundances exceed current *in situ* capabilities [16]. Complex organics are present from the upper atmosphere of Titan to the surface. Sample return enables detailed studies of both the gas and solid components, providing chemical structures and high-precision, compound-specific $^{13}C$ and $^{15}N$ measurements unattainable *in situ* to vastly advance our understanding of Titan abiotic chemistry. At both Titan and Venus, access to cloud particles provides a connection to understanding the planetary volatile cycles (sulfur for Venus, hydrocarbons for Titan). ***Tenuous***

***Atmospheres.*** **Mars** sample return [18] should provide the opportunity to collect sufficient atmospheric samples to enable high-precision isotopic measurements of trace gas species not possible with *in situ* technology, isolated from exchange with drilled or scooped surface material. Additionally, this allows for regolith-atmosphere interactions to be better constrained.

***Astrophysics Value*:** The locations and abundances of volatiles in a stellar system are key to understanding planet formation and evolution, interpreting exoplanet observations, and informing planet formation theory. The > 4100 confirmed exoplanets span a broad range of types and orbital periods which encode the processes of planet formation and evolution, including the bulk locations and distributions of volatiles. Gas and dust disks associated with planet formation exhibit ring, gap, and spiral features [19] which may indicate volatile ice-lines ($H_2O$, CO, $CO_2$, $N_2$, etc.) and planet formation [20]. These systems are starting to provide direct links between exoplanet observation and planet formation and evolution theory. Detailed sample return studies of volatiles in the Solar System will inform the interpretation of exoplanet and disk observations and planet formation and evolution around other stars.

## 3. Technology Development

Key science requirements drive a number of technological considerations, primarily in the thermal and mechanical areas. For example, the sample must be kept sufficiently cold to prevent alteration (e.g., for water: aqueous at $T \leq 263$ K; amorphous-crystalline transition near ~135 K) throughout the entire mission. The sample must also be hermetically sealed before re-entry into Earth's atmosphere to avoid volatile loss and contamination, and the integrity of the seal must be verified *in situ* and maintained through sample analysis in a terrestrial laboratory. *In situ* sample containment and seal verification will be particularly important for Planetary Protection Category V restricted Earth return missions from currently habitable environments (e.g., ocean world plume return) that could host extant life. Initial characterization of the sample at the time of collection to constrain potential volatile loss and/or sample contamination during collection and documentation of the sample temperature and pressure history from collection through curation will also be critical for understanding any changes to the chemical composition of the sample after collection and containment. To minimize volatile contamination of the sampling site from spacecraft thrusters during collection, investments in high-thrust, high-precision clean propellant technologies should be made.

***Collection.*** Technologies that have been developed for collection from hard icy surfaces include drills (Mars rovers), cold directed gas (*OSIRIS-REx*, *CAESAR*), BiBlade (*CONDOR*)[21], and hollow impactor (*Harpoon*[1]). Drills can be used on very hard surfaces but can cause significant heating of the material due to friction. Cold directed gas is limited to loosely packed granular surfaces, so the target object will need to be well understood. The hollow impactor has the potential for being used to collect samples from a variety of surface densities with a minimal increase in temperature of the collected material during collection. The hollow impactor has the added benefit of sampling below the surface away from potential spacecraft contaminants. See white paper by Westphal et al. and Bockelee-Morvan (*AMBITION*) for details on cryogenic sample return.

Storing the volatile sample after collection may require the separation of the refractory components from volatiles. Processes for separating water-ice from refractory samples have been demonstrated (*CAESAR*), however colder methods need to be developed.

[1] *Work done by J. Nuth, D. Wegel, L. Purves, E. Amatucci, M. Amato (GSFC). GSC-16869-1*

***Temperature control during entry, descent, and landing (EDL).*** Technologies developed to control temperature during re-entry and recovery include the ability to jettison the heat shields after re-entry during descent, before the heat of re-entry significantly increases the thermal environment surrounding the sample such as done with *Hayabusa* [22]. Hydrocarbon-based phase-changing materials (PCM) have been proposed as a means of maintaining low-temperature storage during EDL and recovery at -10°C and could be feasible for maintaining temperatures from 0°C to -100°C long enough for EDL requirements. Maintaining temperatures below -100°C would take significant improvement in existing technologies such as low power-consumption cryocoolers or extremely short EDL mission segments.

## 4. Laboratory Studies

Volatile sample return missions must consider the potential for sample alteration from collection to return to curation. Warming could induce physical changes, chemical reactions, and isotopic fractionation. If altered, laboratory simulations can deduce some of the original nature of returned samples. However, those simulations rely on details of the history of the sample temperature and pressure throughout the entire mission; the lower the temperature excursion and the higher-fidelity the record, the greater the science return. **Finding: To extract the maximum amount of science impact, laboratory studies to provide a threshold requirement need to be supported for any sample return mission.**

Laboratory experiments concomitant with mission development should focus on documenting the limits of environmental conditions that preserve the science and to identify sample preservation strategies against temperature, pressure, isotope exchange or chemical reactions with purge gases and flight hardware, and contamination. Experiments should be performed again with actual measured conditions from collection through curation.

## 5. Curation and Long-term Storage

The curation of volatiles focuses on preserving the sample after Earth return and facilitating sample science to the greatest extent possible. Some volatiles may be immediately analyzed upon return. For longer-term preservation, special conditions and operations are required. Preservation of the physical state of the sample will need to be defined based on the mission science requirements: solid and gas sample storage have different temperature and handling conditions. If the volatiles are mixed with the solid portion of the sample, colder storage temperatures will be needed to slow or prevent chemical reactions. Sample temperature requirements and species of interest also dictate the storage and purge. In all cases, the sample must be isolated from Earth's atmosphere to minimize contamination from terrestrial gases. Storage materials must be compatible to avoid reactions with typical curatorial materials (e.g. steel). Preliminary examination (PE) to document the sample for cataloguing typically includes weighing, photographing, and nondestructive classification of the sample. For gases and ices, PE will need to allow for rapid analysis of samples without contamination or alteration while consuming the lowest amount of sample possible. After PE, the sample would be placed in cold storage and monitored to ensure sample integrity for the long term.

**Finding: Develop technologies, procedures, and operations plans for limiting volatile-volatile and/or volatile-solid chemical reactions during return. Finding: Investigate the safety hazards of working with toxic/hazardous materials, and implement mitigations to those hazards as part of mission planning.** Materials compatibility, toxic chemical monitoring, sample containment & handling; and sample handling procedures may be more complicated. **Finding:**

**Develop curatorial technologies, infrastructure, and procedures for minimizing chemical alteration of volatile-bearing samples during long-term storage.**

Long-term storage procedures must minimize the possibility of chemical changes to the sample. Analyses of *Stardust* Al foils showed that volatile precursors to glycine were lost over months of curation [12]. Mitigation approaches such as sealing and leak rate monitoring can address both contamination and losses. Keeping the returned samples cold mitigates volatile loss, but could enhance trapping of contaminants. **Finding: New PE procedures must be developed to handle gas and ice samples.**

***Table 1. Science enabled by different temperature thresholds.*** *Science objectives are from questions: "How and where do volatile materials - including water - that shape the bulk, surface, atmosphere, and habitability of planetary bodies form?"; "What is the origin of* ***organic compounds****?"; "How do noble gases record object accretion?" Temperatures are conservative as they apply to the pure material. Species trapped in solid matrices may be captured at higher temperatures. Temperature-independent samples are unreactive noble gases and $N_2$.*

| Science Question | Measurement/ Sample Volume | Temp* | Enabling Technology | Collection Consideration | Key Species |
|---|---|---|---|---|---|
| Organic formation pathways | Organics > 100 Da<br>· 1 nmol of analyte | 272 K | PCM ($NaCl/H_2O$ eutectic, dodecane) | Volatile isolation from solid material | $H_2O$, >$C_8$ organics |
| C, N, & S formation pathways | *As above, plus:*<br>· Stable isotopes for most C, N, & S unreacted volatiles<br>· 2 nmol analyte C<br>· 3 nmol analyte N & S | 165 K cryo | $CH_3OH$/ $H_2O$ eutectic | *As above, plus:*<br>· isotope exchange<br>· Sample container materials | *As above, plus:* Xe, all organics but $CH_4$ - $C_3H_8$, $CO_2$, $NH_3$, $NO_2$, $H_2S$, $SO_2$, HCl |
| C and N formation pathways, ice physical structures | *As above, plus:*<br>· Pristine organic composition<br>· Isotopic measurements on all C, N unreacted volatiles<br>· $H_2O$ ice phase (amorphous vs. crystalline, clathrates)<br>· 1 $\mu m^2$ ice film for ice structure analysis | 77 K cryo | Active cryocoolers, $LN_2$ | *As above, plus:*<br>· backfill cryo-chamber?<br>· ice-phase preservation during collection<br>· cryopumping sample onto substrate?<br>· radiation isolation | *As above, plus:* Ar, Kr, all organics, CO, $CH_4$, NO |
| Formation pathways of H and O compounds | *As above, plus*<br>· isotopes of all H, O unreacted volatiles<br>· 20 nmol analyte for H | 3-14 K cryo | LHe $SH_2$ | *As above, plus:*<br>· inert gas backfill?<br>· H exchange reactions | *As above, plus:* Ne, $H_2$, $N_2$, $O_2$ |

*Maximum temperature of sample during collection, entry, descent, landing, transport, and curation.

## 6. Planetary Protection and Contamination Control

Sample return targets such as the Moon and small bodies are expected to be Category V unrestricted Earth return due to the history of radiation, lack of liquid water, etc. However, sample return from objects with potential extant biology, such as Mars and ocean worlds, would be expected to be Category V restricted Earth return [12]. Particular attention to avoiding volatile loss and fractionation must be applied to volatile samples. Heat sterilization, though effective, would drive reactions and destroy ices. Other methods of sterilization from categorized sites that may have an acceptable level of alteration of the sample (e.g. ionizing radiation (x-ray, ɣ, $^{0}$n or β particles)) can be employed. Since the sample must be returned hermetically sealed to prevent air contamination, the ability to chemically sterilize the container exterior via strong oxidizing and/or reducing agents is possible. Leveraging propellant (e.g. $N_2H_4$ and $N_2O_4$) plumbing technologies could be effective. Current forward planetary protection cleaning and sterilization technologies, such as dry heat microbial reduction (DHMR) should largely be sufficient, though more heat-tolerant materials are desirable. **Finding: New technologies for in-flight sterilization and encapsulation of restricted materials need to be developed and tested.**

However, for all sample return missions contamination control and knowledge are particularly important [23]. Terrestrial contamination (including sterilized microbes) is a serious impediment to analysis of sample return material. **Finding: Volatile sample return missions must control additional sources of contamination with a viable pathway to the sample container.** Thruster plume products (e.g. Xe, $N_2$, $H_2$, $NH_3$), spacecraft outgassing (e.g. $H_2O$, pyrotechnic fastener products), and air from assembly through launch as well as Earth return, recovery, and curation (e.g. $H_2O$, $CH_4$, $H_2S$, $N_2$, $O_2$, $CO_2$, CO, Ar) must be controlled, depending on the specific sample return temperature ranges and science targets.

## 7. References

[1] Bernstein MP et al. (1995) ApJ 454, 327; Matthews CN Minard RD (2006) Faraday Discuss 133, 393; Cottin H et al. (2001) ApJ 561, L139; Balsiger H et al. (2007) Space Sci Rev 128, 745; Gulkis S et al. (2007) Space Sci Rev 128, 561; Hässig M et al. (2013) Planet Space Sci 84, 148; Lee S et al. (2014) AAS/DPS 46, id.100.05; Altwegg K et al. (2015) Science 347, 6220; de Marcellus P et al. (2015) PNAS 112, 965; Goesmann F et al. (2015) Science 349, 6247; Hässig M et al. (2015) Planet and Space Sci 105, 175; Le Roy L et al. (2015) A&A 583, A1; [2] Liley PE, et al. (1997) Perry's Chemical Engineer's Handbook 2, 374; [3] Ahrens et al. white paper ; [4] De Sanctis MC et al. (2015) Nature 528, 241; Marchi S et al. (2019) Nature Astro 3, 140; [5] Combe JP (2016) Science, 353, p.aaf3010; [6] Kaplan HH et al. (2018) Geophys Res Let 45, 5274; [7] Castillo-Rogez JC et al. (2020). Astrobio 20, 269; Castillo-Rogez JC et al. white paper; [8] Spencer JR et al. (1997) Geoph Res Let 24, 2471; [9] Porco CC et al. (2017) Astrobio 17, 876; [10] Soderblom LA, et al. (1990) Science 250, 410; [11] Tsou P et al. (2012) Astrobio 12, 730; [12] Neveu M et al. white paper; [13] Baines KH et al. (2013). In Comparative Climatology of Terrestrial Planets, Mackwell SJ et al., eds.), 137; [14] Glaze L et al. (2016), LPSC 1560; [15] Rabinovitch J et al. (2019) VEXAG 8009; [16] Glein CR (2017) Icarus, 293, 231; [17] Greenwood RC Anand M (2020) Space Sci Rev, 216, 52; [18] e.g. McSween white paper; [19] Boccaletti A et al. (2020) A&A 637, L5; Huang J et al. (2018) ApJ 852, 2; [20] Keppler M et al. (2018) A&A 617, 44; [21] Backes P et al. IEEE Aerospace Conference, March 2017. [22] Yamada et al. (2012) Trans. JSASS 10 Te_11; [23] Dworkin JP et al. (2018) Space Sci Rev 214, 19; Chan QHS et al. (2020) Space Sci Rev, 216, 56.